# Daylight entanglement-based quantum key distribution with a quantum dot source

F. Basso Basset[1,*], M. Valeri[1,*], J. Neuwirth[1], E. Polino[1], M. B. Rota[1], D. Poderini[1], C. Pardo[1], G. Rodari[1], E. Roccia[1], S. F. Covre da Silva[2], G. Ronco[1], N. Spagnolo[1], A. Rastelli[2], G. Carvacho[1], F. Sciarrino[1,†], and R. Trotta[1,†]

[1]Department of Physics, Sapienza University of Rome, 00185 Rome, Italy
[2]Institute of Semiconductor and Solid State Physics, Johannes Kepler University, 4040 Linz, Austria

[†]rinaldo.trotta@uniroma1.it, fabio.sciarrino@uniroma1.it
[*]These authors contributed equally.

**Entanglement-based quantum key distribution can enable secure communication in trusted node-free networks and over long distances. Although implementations exist both in fiber and in free space, the latter approach is often considered challenging due to environmental factors. Here, we implement a quantum communication protocol during daytime for the first time using a quantum dot source. This technology presents advantages in terms of narrower spectral bandwidth—beneficial for filtering out sunlight—and negligible multiphoton emission at peak brightness. We demonstrate continuous operation over the course of three and a half days, across an urban 270-m-long free-space optical link, under different light and weather conditions.**

## INTRODUCTION

Entanglement is a fundamental resource in quantum networks, enabling functionalities that go beyond those unlocked by prepare-and-measure approaches[1]. In quantum key distribution (QKD)—arguably one of the most established applications of quantum technologies—entanglement-based protocols can offer some advantages compared to more straightforward implementations based on the exchange of single photons. These protocols inherently provide source-independent security[2], a property which has been used to share secret keys without the need for trusted nodes in notable test cases, ranging from free space optical links[3], to metropolitan networks[4] and satellite-based communication over distances of 1120 km[5]. Entanglement-based QKD is also naturally implemented in quantum repeater architectures[6], which are being developed to overcome current distance limitations in quantum communication. Finally, it can provide additional security towards certain individual eavesdropping attacks[7,8] and potentially make possible the demonstration of fully device-independent operations[9].

Depending on the targeted application, QKD protocols can be implemented either in optical-fiber or free-space communication channels. While telecom wavelengths are favorable for long-distance fiber transmission, visible or short near-infrared can be advantageous in free space. This is argued for satellite links[10], mainly due to aperture-matching losses related to diffraction, in addition to the compatibility with practical silicon avalanche photodiode detectors. The advantage can persist even in ground-to-ground links and during daytime operation, despite background noise from sunlight and turbulence would favor telecom wavelengths[11].

Most free-space quantum communication experiments have been demonstrated under favorable conditions, such as absence of sunlight and good weather conditions, often for short periods of time, in order to avoid drops in performance related to high background noise and misalignments of optical systems. However, addressing these problems is crucial to ensure real-life quantum communications based on long-distance free-space channels. This is the case for ground-to-satellite links, which require continuous operations in different weather conditions. Previous works[12–18] have achieved this task by adopting spatial, temporal and wavelength filters, or by using coincident event-based protocols to reduce background noise. Active feedback systems have been used to stabilize the quantum channel over time in different weather conditions, even using drones[19]. QKD during daylight has been extensively demonstrated using prepare-and-measure protocols[12–16] and is an active research direction for solutions based on continuous-variable encoding[17]. However, only a single experimental study has tackled this challenge for entanglement-based QKD[18], using a spontaneous parametric down conversion (SPDC) source in combination with an interference filter with 6.7 nm bandwidth to narrow down a signal with spectral width of 8.7 nm. Most common SPDC sources suffer from large bandwidths, preventing the use of narrower wavelength filters to reject background photons that lower the efficiency and fidelity of the protocol. In fact, this has been considered a major hurdle against the development of entanglement-based protocols under daylight conditions[11]. A viable solution is represented by quantum dot (QD) single-photon sources, as discussed below.

As other quantum emitters, QDs allow overcoming the trade-off between brightness and multiphoton generation which limits the performance of Poissonian sources. QDs have been used to perform prepare-and-measure protocols, starting from first demonstrations of a performance edge with respect to weak coherent pulses in the standard BB84 protocol[20–22] to more recent studies which foresee an advantage over decoy-state protocols as well[23,24]. While developing a deterministic entangled photon source is crucial to offset distance limitations linked to the signal-to-noise ratio[25], QDs also have an advantage in terms of spectral bandwidth. In fact, their emission spectrum is compatible with ultra-narrow filters, such as the volume Bragg grating with 0.05 nm full-width at half-maximum (FWHM) used to enable decoy-state QKD over a ground-to-ground distance of 53 km in daylight[14].

Entanglement-based QKD has been realized with QDs, first in-lab[26] and recently in urban links[27,28]. Apart from QKD, other quantum communication protocols have been investigated using entangled photons from QDs, such as quantum teleportation[29,30] and hybrid quantum networks[31,32]. Despite supporting secret key exchange across a free-space optical channel[27], the use of QDs in daylight conditions has not been explored yet, not even for prepare-and-measure protocols[33].

In this work, we experimentally implement entanglement-based QKD in daylight using a QD-based photon source. We implement an Ekert-like protocol[34,35] using a QD under resonant two-photon excitation, which enables the deterministic generation of entangled photon pairs. We discuss the design elements of the free space channel which are introduced to minimize the impact of environmental conditions. With our implementation we establish secure communication during continuous, alignment-free and three-day long operation, showing minimal variability in performances. In particular, we are able to exchange 106 bps averaged key rate during day, night and rainy environmental conditions, while achieving a corresponding secure key rate of 12 bps. Thus, we demonstrate that a QD-based source can support entanglement-based QKD, with the capability to properly operate in an urban environment and showing significant robustness for the free-space QKD optical link in daylight.

# RESULTS

## PROTOCOL

In our realization of entanglement-based QKD, we opted for a protocol derived from Ekert91[34], in which eavesdropping attempts are first detected via the experimental test of a Bell inequality rather than via parameter estimation of the quantum bit error rate (QBER) on a subset of the shared key. More specifically, we used the protocol proposed by A. Acín and coworkers, which reduces the number of measurement bases and can be extended to device-independent operations[9].

Two parties, Alice and Bob, receive one photon each from an entangled pair and perform a measurement on a basis randomly chosen from a given set. For polarization entangled photons prepared in $|\phi^+\rangle = 1/\sqrt{2}(|HH\rangle + |VV\rangle)$, the sets of bases are $\{A_k, A_0, A_1\}$ = {H/V, -22.5°/67.5°, -67.5°/22.5°} and $\{B_0, B_1\}$ = {H/V, -45°/45°}, respectively for Alice and Bob (H/V indicate horizontal and vertical linear polarization in the laboratory frame of reference, other linear polarizations are indicated with the angle relative to H). After the measurement bases are communicated, when Alice and Bob have performed the same measurement ($\{A_k, B_0\}$), they can add the result as a bit to a shared secret key.

At this stage (sifting), the two keys have discrepancies quantified by the QBER:

$$Q = (1 - E(A_k, B_0))/2 \qquad (1)$$

where $E(A_i, B_j)$ is the correlation coefficient. This is the expectation value on the given pair of measurements, which is calculated from the coincidences $n_{i,j}$ recorded for all the possible combinations of results $\{i, \overline{i}\}$ and $\{j, \overline{j}\}$:

$$E(A_i, B_j) = \frac{n_{i,j} + n_{\overline{i},\overline{j}} - n_{\overline{i},j} - n_{i,\overline{j}}}{n_{i,j} + n_{\overline{i},\overline{j}} + n_{\overline{i},j} + n_{i,\overline{j}}} \qquad (2)$$

If a combination of bases $\{A_0, A_1\}$ and $\{B_0, B_1\}$ is recorded, the output can be used to check the violation of the Bell inequality $|S| \leq 2$, where the parameter $S$ is defined according to the Clauser–Horne–Shimony–Holt (CHSH) definition[36]:

$$S = E(A_0, B_0) + E(A_0, B_1) - E(A_1, B_0) + E(A_1, B_1) \qquad (3)$$

The CHSH parameter acts as a first gauge for eavesdropping attempts. Then, in conjunction with the QBER estimation required during the error correction stage, the CHSH parameter is used in the privacy amplification step to determine the size of the secret key which is finally distilled using a hash function[37].

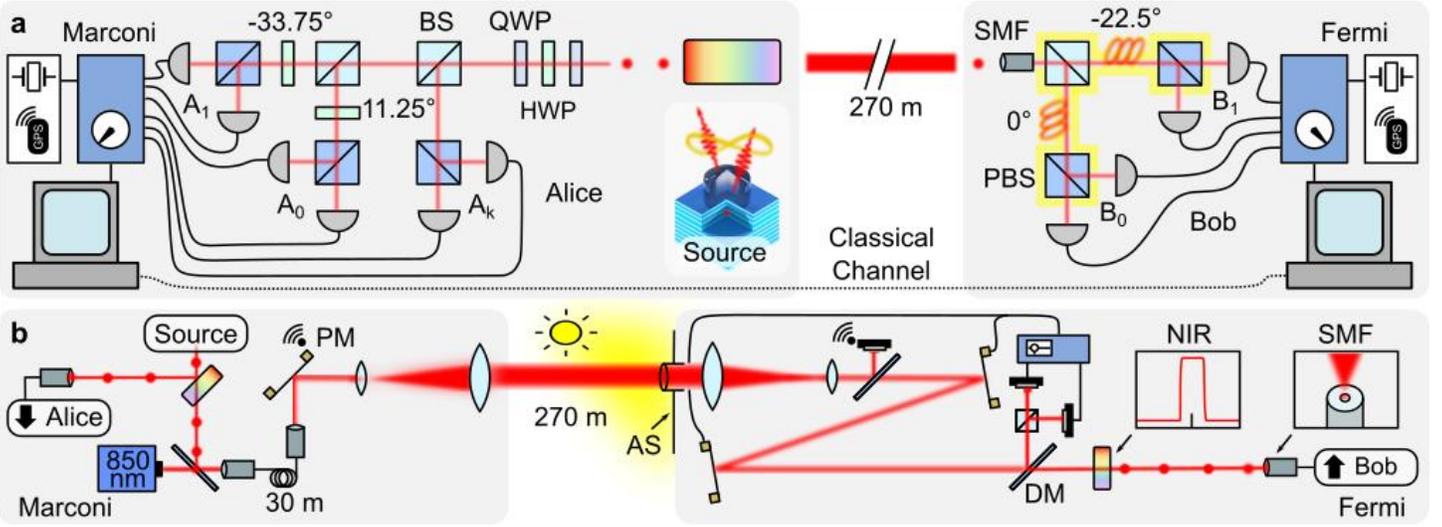

**Fig. 1 QKD setup. a** Two polarization entangled photons generated by a QD are spectrally separated, and one is sent over the free-space channel. After a polarization compensation with a set of two quarter-wave plates (QWP) and a half-wave plate (HWP), the photons are randomly split by 50:50 beam splitters (BS) and measured in different polarization bases using a HWP or fiber paddles and a polarizing beam splitter (PBS). The signal recorded from avalanche photodiodes is collected by independent time-to-digital converters, which are synchronized with the help of GPS-disciplined oscillators. **b** Free-space channel setup. A laser beacon is sent together with the quantum signal over 270 m and separated by dichroic mirrors (DM) at the receiver. It feeds two closed-loop stabilization systems, for fast and slow drift correction respectively, which drive mirrors with piezoelectric adjusters (PM). Background light is minimized by an aperture stop (AS), a 3 nm-FWHM narrow spectral filter (NIR), and a single-mode fiber (SMF).

## SOURCE AND DETECTION

Figure 1 shows the experimental implementation of the QKD protocol. Pairs of entangled photons are generated from a single GaAs/AlGaAs QD (see Methods). This class of quantum emitters has been used to demonstrate extremely low values of multiphoton emission[38] and fidelities up to 0.98 to a Bell state without the need for temporal or spectral filtering[28,39]. We drive the QD under resonant two-photon excitation, a process which allows for the deterministic generation of polarization-entangled photon pairs[40], with a repetition rate of 320 MHz, compatible with the short lifetime of these emitters[41]. The device throughput is therefore mainly limited by the single-photon extraction efficiency. For the QD used in this work it is estimated equal to 10%, using a sample design already employed for demonstrations of QKD[27,28], as well as in more demanding 4-photon protocols[42]. State-of-the-art devices have proven extraction efficiencies up to 85%[43,44] at about 0.90 entanglement fidelity though with the aid of nanophotonic cavities. We select a QD for which, in the laboratory, we measure a fidelity to a maximally entangled Bell state of 0.942(13), a violation of the Bell inequality with the parameter $S$ = 2.606(9), and a visibility in the H/V basis that would correspond to a QBER = 3.2(3)%. We also observed $g^{(2)}$ values of 0.013(1) and 0.022(2) for the two QD emission lines, respectively, proving the expected single-photon behavior.

Figure 1a illustrates how the basis selection and measurements are implemented in the optical setup (see Methods for further detail). One photon of the pair is kept in the laboratory, where Alice performs measurements in the set of bases $\{A_k, A_0, A_1\}$, while the other is transmitted along a 270-m-long free-space channel and collected by Bob who uses the bases $\{B_0, B_1\}$. The photons are detected by silicon avalanche photodiodes, each associated to a specific polarization state. The detection events are recorded by time-to-digital converters, synchronized with the help of GPS-disciplined oscillators.

Note that, along the free-space channel photons are transmitted at 784.75 nm, corresponding to the biexciton-to-exciton transition from the QD, within one of the common wavelength windows for free-space optical communication. Their spectral bandwidth is narrower than the spectral resolution of our spectrometer, that is approximately 0.02 nm (40 μeV). This number is already compatible with the aforementioned noise-suppression strategies based on extra-narrow spectral filtering, a clear advantage compared to approaches that use SPDC.

## FULL-DAY OPERATION FREE-SPACE CHANNEL

Alice and Bob are located in two different buildings of the main campus of the Sapienza University of Rome, separated by a distance of about 270 m. Figure 1b presents the transmitter and receiver setup used to transfer the optical signal. The quantum signal, together with a reference laser, is expanded in a transmission telescope to maintain collimation. At the receiver, the beam is reduced again and redirected by a 200 Hz closed-loop tip-tilt stabilization system to compensate for beam wandering due to air turbulence and mechanical vibrations.

To guarantee operation in daytime, three solutions have been adopted to minimize background noise due to sunlight. First, the access to the receiver platform is only allowed through a clear aperture, whose dimensions are close to the diameter of the first collecting optical element (4 inches), to reduce noise from diffused light. Second, a spectral filter with a bandwidth of 3 nm (FWHM) is used to filter out wavelengths clearly different from the quantum signal. We have measured a 97% transmission on the QD emission line. As previously mentioned much narrower filters are compatible with our source, yet we will show that further filtering is not required in our test case. Finally, the signal is coupled into a single-mode fiber, which acts as a very efficient spatial filter, eliminating light impinging with angles departing from the direct link line.

This is obtained with the active beam stabilization system, which can compensate for most of the instabilities introduced by air turbulence. For the link distance investigated here, these are dominated by the tip-tilt components. We observe a varying single-mode coupling efficiency, approximately from 30% to 50%.

Using this spectral and spatial filtering approach, we measured a background count rate due to sunlight (at 15:00, solar irradiance of 120 W/m$^2$ recorded from the weather station, sun position facing the entrance of the receiver) below 520 counts per second (cps), averaged over all the detectors at the receiver. Since only coincidence events are relevant in Eq. (2) and not single-channel counts, we will see that this level of noise can be easily tolerated.

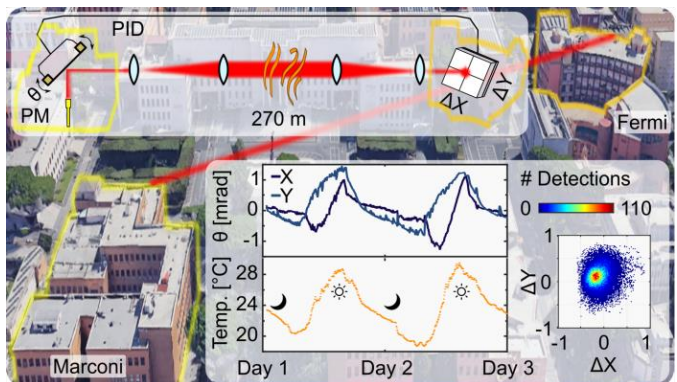

**Fig. 2 Active pointing control.** A PID controller corrects for slow pointing drifts using a PM with feedback from a 4-quadrant detector at the receiver. The variation of outdoor temperature (orange) over two days of testing (x-axis ticks at 12:00 AM) is shown together with the angular motion θ of the PM along tip and tilt directions (blue). The reference laser is maintained on the 4-quadrant detector as shown by the histogram of its position readings (right inset).

As operation during daytime is enabled, continuously running the QKD system becomes possible. In order to achieve this goal, it is important to add elements to guarantee stability over the communication channel. Specifically, we introduce a further pointing system with active feedback at the transmitter. While it has been argued that this approach is not essential for distances below 2.5 km under an average turbulent regime[45], we observe that it is still required for long-term stability due to slow drifts in the transmitter pointing direction.

The concept of the pointing system is depicted in Fig. 2 in the context of the urban link where the key exchange is performed. A mirror with piezoelectric adjusters (PM) is driven by a custom proportional–integral–derivative controller (PID) based on the readings of a 4-quadrant detector (CCD) placed at the receiver. The laser beacon is partially reflected and redirected onto the CCD by a dichroic mirror which entirely transmits the QD signal instead. Each step of the feedback loop takes 10 s, which is sufficient to average over faster beam wandering due to turbulence. The angular motion of the mirror in the tip-tilt directions is reported for a test period of two days. Variations in outdoor temperature are also shown and appear to be correlated with the drifts in pointing direction, which could suggest their attribution to thermal effects on the optical elements of the transmitter. During the test, the beam always falls entirely within the 4-quadrant detector as shown by the histogram of readings in Fig. 2. The position distribution is not exactly centered due to some degree of wind-up, which is however compensated for by the receiver stabilization system: indeed, the beam fully stays within the aperture of the collection optics and the coupling with the single-mode fiber is maintained.

**EXPERIMENTAL RESULTS**

After having established the optical link, we performed entanglement-based QKD for a period of three days and a half. The data were collected between November 12$^{th}$ and November 15$^{th}$ 2021, during day and night and experiencing different environmental conditions such as clear and cloudy sky and in the presence of rain. Figure 3 reports the main figures of merit related to the performance of the QKD protocol. On average we observe a sifted key rate of 106 bit/s, with a QBER of 7.16(2)% and a CHSH parameter of 2.409(2). For immediate comparison, we also report the values of solar irradiance $I_{sun}$, which depend on the hour of the day and on the cloud cover, and the rain rate $R_{rain}$ (if present), which are the main weather variables with the potential to impact the operation of the optical link.

Despite the varying environmental conditions, we demonstrate that the figures of merit remain rather stable and well within the thresholds of the QKD protocol. The sifted key rate does not show any significant correlation with the level of sunlight. In fact, the dominant effect is due to oscillations in the coupling efficiency at the receiver. Despite the uninterrupted connection, there are still signal fluctuations—on the timescale of hours—amounting to a root-mean-square deviation of 17% of the average single-photon count rate at the receiver. These can be mainly attributed to air turbulence, in which case an additional adaptive optics kit could be employed to further improve stability[46,47].

On the contrary, the QBER (CHSH parameter) does slightly increase (decrease) with higher level of sunlight. We calculated that this behavior is consistent with the level of background detection events due to incomplete filtering of the external light. To do so, we experimentally measured the daylight accidental coincidences $n^{acc}$ between the QD signal on Bob's side and background events only on Alice's end, by blocking the single-photon signal from entering the sender platform at a near-average value of solar radiation (120 W/m$^2$). We then add them to the coincidences recorded during night ($n^{night}$), as

$$n_{i,j}^{day}(I_{sun}) = n_{i,j}^{night} + n_{i,j}^{acc}(120\,W/m^2)\frac{I_{sun}}{120\,W/m^2} \quad (4)$$

and substitute these figures in Eq. (2) to simulate the relevant figures of merit during daytime. This procedure returns increases in QBER by 0.013 and decreases in $S$ by 0.06 for a solar irradiation of 500 W/m$^2$, which are consistent with the maximal variations observed.

It is worth noticing that the signal-to-noise ratio on the coincidences $n_{sig}/n_{acc}$ is in general higher than the one on the single photon detection events $R_{sig}/R_{acc}$. More specifically, it depends on the coincidence time window with respect to the period between two pump pulses $\Delta t_{coinc}/T_{rep}$ and on the probability of photon pair generation $p_{pair}$ as follows:

$$\frac{n_{sig}}{n_{acc}} = \frac{R_{sig}}{R_{acc}}\left(p_{pair}\frac{\Delta t_{coinc}}{T_{rep}/2}\right)^{-1} \quad (5)$$

In our case, the coincidence time window of 1.3 ns causes a factor 1.2 increase, whereas $p_{pair}$ accounts for an additional factor 4.5. This is due to the presence of blinking, despite the use of highly-efficient resonant two-photon excitation[40] (see figures in Methods). Equation (5) also allows us to estimate the daytime maximal variation of CHSH parameter and QBER knowing the background and signal individual count rates.

In the presence of rain, we observe no significant impact on CHSH parameter and QBER. While there is a concomitant decrease in the detected events at the receiver, the signal-to-noise does not decrease as much as in the case of sunlight, at least for the investigated levels of rainfall with peaks of 20 mm/hr.

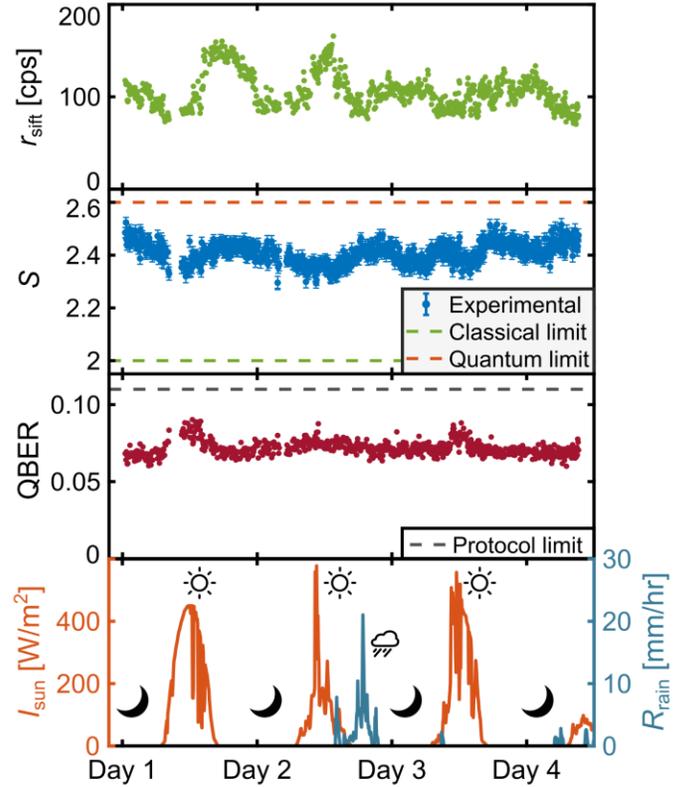

**Fig. 3 QKD performance for day-and-night operation.** The sifted key rate (green), CHSH parameter (blue), QBER (red) measured over three and a half days of continuous operation. The error bars in the CHSH parameter are calculated by Gaussian propagation assuming a Poissonian distribution for the coincidence counts. Each data point is averaged over 100 acquisitions of 1.2 s. Relevant weather data, recorded from a nearby meteorological station (CNR Sede Centrale, distant 300 m from the receiver), are reported over the same time span.

The QBER is ultimately limited here by other factors, namely a source with a slightly lower level of entanglement compared to previous work[27] and, arguably, a depolarization effect in the free-space channel implementation. Both these shortcomings are not fundamental in nature.

With the achieved figures of merit, we were able to exchange a secret key over the whole duration of the transmission. Considering the protocol we use, the secure key rate $r_{sec}$ can be estimated in the asymptotic limit[48] starting from the sifted key rate $r_{sift}$ with:

$$r_{sec} = r_{sift}\left(1 - f_{EC}h(Q) - h\left(Q + \frac{S}{2\sqrt{2}}\right)\right) \quad (6)$$

where $h(x) = -x \cdot log(x) - (1-x) \cdot log(1-x)$ is the binary Shannon entropy, and $f_{EC}$ is the efficiency[49] of the error correction algorithm. In our case, we adopted the CASCADE algorithm[49,50], commonly used in modern implementation of QKD protocols, which in this case has a theoretical efficiency[51] of $f_{EC} = 1.2$.

In our implementation, Alice and Bob continuously record coincidences in 6 s frames. Following the protocol, they can compute the actual CHSH value from coincidences between the basis {$A_0$, $A_1$} and {$B_0$, $B_1$}; if $S < 2$, the corresponding frame is discarded, since security cannot be guaranteed. Otherwise, the frame is accepted, and the sifted key is computed on the events ($A_k$, $B_0$). During the duration of the experiment, we were able to exchange a total of 1305 kB (~106 bit/s).

To properly perform the error correction, the CASCADE algorithm requires as an input the QBER of the frame, which has been estimated by announcing (and thus discarding) a fixed fraction of the sifted key, namely 30%. Overall, prior to error correction, we have at our disposal 915 kB of key data. During error correction, Alice and Bob exchange information on a classical channel, resulting in a total error corrected key of 495 kB (~40 bit/s) and an actual efficiency $\hat{f}_{EC} = 1.26$.

To compute the final secret key rate, note that the last term in Eq. (6) is related to the information leaked to a hypothetical eavesdropper, which has to be minimized by the privacy amplification procedure. Here, for each frame, it has been computed using both the measured CHSH parameter $S$ and the estimated QBER, using the most commonly employed security bound[9]. If the rate is negative, or the estimated QBER is greater than 11%, no secret key can be extracted, and the corresponding frame is discarded altogether. At the cost of further reducing the secret key rate, the protocol is also compatible with a stricter bound only relying on the CHSH parameter estimation in case one wants to reduce the assumptions on the measurement devices. To computationally perform the privacy amplification step and extract the final secret key, we employ the Trevisan extractor, a construction for universal hash functions[37] which has been shown to be secure also in the presence of quantum side information[52,53]. The achieved length of the final secure key is 142 kB (~11.5 bit/s).

## DISCUSSION

The receiver design that we employed to reduce the impact of background light on the performance of the protocol also has the benefit of preventing specific eavesdropping strategies. More specifically, spatial selection via a single-mode fiber prevents spatial mode side-channels attacks[54], while the use of a spectral filter with a bandwidth narrower in wavelength than the range of nominal operation of the 50:50 beam splitter counters frequency-dependent beam-splitting ratio attacks[55]. Photon number-splitting and beam-splitting attacks are not relevant due to the anti-bunched nature of the QD source and the use of an entanglement-based protocol[7]. On the other hand, it is outside of the scope of the article to explore all viable side-channel attacks, including approaches targeting known detector flaws[56–58], more general efficiency mismatches[59,60], and finite-key size analysis[61].

In conclusion, we have demonstrated that a QD photon source can support entanglement-based QKD over a 270 m free-space channel continuatively for a total of 82 hours, therefore withstanding daytime—and even moderate rain—conditions. This is a crucial requirement for the use of such technology in practical scenarios. We achieved a secure key rate of 12 bps during daylight and nightlight conditions, without significant variations of QBER. Moreover, we quantified the amount of residual accidental events due to background light and offered a simple criterion to estimate its impact on the performance of the protocol. In fact, the characteristics of the QD-based entangled photon source grant sufficient headroom to further extend operation to considerably longer distances. Turbulence-related effects can be tackled by an adaptive optics system, with the 780 nm wavelength benefiting from lower beam diffraction during propagation compared to light at telecom wavelengths[11]. With respect to environmental noise, the bandwidth of the source, which here is less than 0.02 nm (down to 0.003 nm at the Fourier limit), can be reduced by two to three orders of magnitude to allow for a larger telescope aperture while maintaining comparable noise levels. Concerning the signal intensity, optimizations in the sender and receiver setup are well in reach and include reduced optical losses and superconducting nanowire detectors, which could improve detection efficiency and temporal selection. More importantly though, applying recent experimental improvements in the design of the source[43,44] would lead to an increase in pair generation probability up to more than two orders of magnitude. At the same time, the fast recombination times are compatible with driving rates above the GHz[62]. By combining these advances with strategies to simultaneously optimize the degree of entanglement[39], we foresee that daytime operation can be achieved with an entanglement based protocol at distances so far only explored by prepare-and-measure QKD schemes[14]. Finally, in order to increase the security and the efficiency of the shared key in a fully black-box approach, suitable optimization algorithms could be used during the operations[63]. These findings will assist the development of more practical free-space communication infrastructures, which will have to play an essential role in the realization of long-distance quantum networks.

## METHODS

### ENTANGLED PHOTON SOURCE

Our source of polarization-entangled photons is a single GaAs/Al$_{0.4}$Ga$_{0.6}$As QD fabricated using droplet etching epitaxy[64]. The growth parameters were optimized to yield nanostructures with maximal in-plane symmetry to reduce the average fine structure splitting of QDs in the sample[65]. The QDs are placed at the middle of a planar distributed Bragg reflection cavity, whose geometry is described in previous works[42], designed to improve the probability of collecting light from the top of the sample. This, in combination with a zirconia Weierstrass solid immersion lens, results in an estimated extraction efficiency of 10% in a 0.5 NA aspheric lens. The selected QD emits entangled photons via the biexciton-exciton (XX-X) cascade. The two emission lines have wavelengths $\lambda_{XX}$ = 784.75 nm and $\lambda_X$ = 782.86 nm and measured zero-time delay intensity autocorrelation $g^{(2)}_{XX}(0) = 0.013(1)$ and $g^{(2)}_X(0) = 0.022(2)$, mainly attributed to the incomplete suppression of the laser light and to the afterpulsing effect of the detectors[66]. The QD was selected with a fine structure splitting of 1.0(5) µeV, resulting in a fidelity to a maximally entangled Bell state of 0.942(13) and a concurrence of 0.88(3), from a 36-bases quantum state tomography performed in the laboratory. Via intensity cross-correlation measurements between XX and X[67], we also estimate the preparation fidelity (probability that a laser pulse induces a radiative cascade from the QD in the ground state) $\eta_{prep} = 0.86$ and the "on"-time fraction related to blinking $\eta_{blink} = 0.26$.

**EXPERIMENTAL SETUP AND PROCEDURES**

The QD source operates at 5 K in a low-vibration closed-cycle He cryostat. It is driven by a Ti:Sapphire femtosecond laser with 80 MHz repetition rate. The rate of pump pulses is increased to 320 MHz using two delay lines, while the spectral linewidth is reduced to 200 µeV for resonant two-photon excitation using a 4f pulse shaper with an adjustable slit on its Fourier plane. The laser is focused on the sample using an aspheric lens with 0.5 NA, the same used to collect the signal. Volume Bragg filters with a spectral bandwidth of 0.4 nm are used both to eliminate laser stray light and to separate the XX and X emission lines. These signals are coupled into separate single-mode fibers. X photons are sent to Alice's analyzer, which is equipped with bulk 50:50 beam splitters for the random basis choice and zero-order half-wave plates plus polarizing beam splitters for the polarization selection (see Fig. 1a). Prior to the analyzer, a set of three wave plates has been used to compensate for setup-induced polarization changes[28,30], so to optimize the CHSH parameter measurement in the laboratory. XX photons are sent to the free-space channel. They are combined with a laser diode at 850 nm, which is used for beam stabilization, and pass through a beam expander. The beam diameter of 22 mm ($1/e^2$ intensity level) allows maintaining collimation for the air travel distance of 270 m. At the receiver, after a beam reducer, the signal is stabilized by an active laser stabilization system (MRC Systems GmbH) consisting of a couple of fast steering mirrors and position sensitive detectors. Its closed-loop frequency is 200 Hz, which sits above most of the turbulence frequency spectrum in normal atmospheric conditions. The overall efficiency of the free-space link is 10% including optical losses on the sender and receiver platforms, attenuation in the free-space channel, and single-mode fiber coupling efficiency. During the key exchange, the average count rate was 470 kcps on the sender's side and 58 kcps on the receiver's side. In addition to the background due to sunlight reported above, the receiver's detectors average 250 cps of dark counts and 700 cps from the 850 nm laser diode. Bob's analyzer is realized with in-fiber optical components, so as to avoid unnecessary additional coupling losses. The polarization bases are calibrated by sending specific linear states of polarized light from the source setup and redirecting them on the corresponding detection channels using paddle (bat-ear) polarization controllers. Further real-time adjustments are not required. The photons are detected by silicon avalanche photodiodes with a time jitter of approximately 400 ps and a nominal detection efficiency of 60% (46% including the receptacle losses), connected to time-to-digital converters with a resolution of 81 ps.

The synchronization between the two parties happens in two steps. First, a coarse common time frame is established by using GPS-disciplined oscillators, both to correct for the internal drift of each time-to-digital converter and to provide a common clock reference with an accuracy around 10–20 ns[33]. Second, a sub-ns accuracy is obtained by analyzing the coincidences on a couple of publicly shared channels, that are not used to generate the secret key, and finding the time delay between correlated detection events in a small range coincidence window of ± 40 ns. The classical information is exchanged with a standard TCP/IP protocol over the university network. Finally, the coincidences are collected within a 1.3 ns acceptance time window.

The execution of the protocol cycles through an acquisition step (1.2 s accumulation time) and a data processing / exchange step, resulting in a duty cycle of 45%. While it was not a major concern in this work, the duty cycle could be improved to 100% simply by software parallelization. In addition to this, almost 30% of the acquisition events were discarded due to unsuccessful synchronization, which is related to irregularities in the reference from the GPS-disciplined oscillators. This technical aspect could be improved by either using reference clocks with better stability or optical signal.

Further details about the free-space optical link budget and the synchronization procedure are reported in previous work[27].

## ACKNOWLEDGMENTS

This work was financially supported by the European Research Council (ERC) under the European Union's Horizon 2020 Research and Innovation Programme (SPQRel, grant agreement no. 679183), and by MIUR (Ministero dell'Istruzione, dell'Università e della Ricerca) via project PRIN 2017 "Taming complexity via QUantum Strategies a Hybrid Integrated Photonic approach" (QUSHIP) Id. 2017SRNBRK. This project has received funding from the European Union's Horizon 2020 Research and Innovation Program under Grant Agreement no. 899814 (Qurope) and 871130 (ASCENT+). A.R. and S.F.C.d.S. acknowledge C. Schimpf for fruitful discussions, the Austrian Science Fund (FWF) projects FG 5, P 30459, I 4320 and the Linz Institute of Technology (LIT).

We thank Salvatore Di Cristofalo for providing us with detailed data from the CNR ENERGY+ weather station in Rome. We thank Radim Filip and Vladyslav Usenko for fruitful discussions.

## AUTHORS' CONTRIBUTIONS

The experiment was performed, in the Marconi building (Alice), by F.B.B., M.B.R., J.N., and C.P., and, in the Fermi building (Bob), by M.V., G.Rod., D.P., and E.P.. D.P. and G.Rod. wrote the software for data acquisition and secret key extraction. J.N., G.Ron., M.B.R., and F.B.B. performed the source characterization. M.V., E.P., N.S., and G.C. designed and assembled the receiver setup and the active stabilization system. F.B.B., E.R., and M.B.R. designed and assembled the source and sender setup. C.P., E.R., and F.B.B. designed the pointing system and tested it with M.V., D.P., G.Rod., and E.P.. S.F.C. and A.R. designed and fabricated the QD-based entangled photon source. F.B.B., J.N., M.V., and G.Rod. wrote the manuscript with feedback from all authors. All the authors participated in the discussion of the results. R.T. and F.S. conceived the experiments and coordinated the project.

## COMPETING INTERESTS

The authors declare no competing interests.